\documentclass[journal=cmatex,manuscript=article]{achemso}

\usepackage[utf8]{inputenc}
\usepackage{graphicx}
\usepackage{mathtools}
\usepackage{subfigure}
\usepackage{amsmath}
\usepackage{amssymb}
\usepackage{xcolor}
\usepackage{amsthm}
\usepackage{csquotes}

\newcommand\sbullet[1][.5]{\mathbin{\vcenter{\hbox{\scalebox{#1}{$\bullet$}}}}}

\title{Excitons  and singlet fission at hybrid inorganic-organic semiconductor interfaces}

\author{M.~V.~Klymenko}
\affiliation{ARC Centre of Excellence in Exciton Science, School of Science, RMIT University, Melbourne, Victoria 3001, Australia}
\email{mike.klymenko@data61.csiro.au}
\author{L.~Z.~Tan} 
\affiliation{The Molecular Foundry, Lawrence Berkeley National Laboratory, Berkeley, California 94720, United States}
\author{S.~P.~Russo}
\affiliation{ARC Centre of Excellence in Exciton Science, School of Science, RMIT University, Melbourne, Victoria 3001, Australia}
\author{J.~H.~Cole}
\affiliation{ARC Centre of Excellence in Exciton Science, School of Science, RMIT University, Melbourne, Victoria 3001, Australia}

\begin{document}

\begin{abstract}
Excitons in organic crystalline semiconductors play a crucial role in the operation of optoelectronic devices such as organic solar cells, light-emitting diodes, and photodetectors. The excitonic properties of materials are dramatically affected by the presence of surfaces and interfaces. In this work, we investigate the influence of a neutral hydrogen-passivated 1x2 reconstructed (100) silicon substrate on excitons within the crystalline tetracene layer deposited on the top of it. Our findings reveal that singlet excitons in the contact tetracene layer are situated within the continuum of unbound Wannier-Mott excitonic states in silicon, with noteworthy hybridization between these states. Consequently, in the contact tetracene layer, all singlet excitons exhibit a pronounced interlayer charge transfer character, while the triplet exciton remains confined to the tetracene layer. This makes the singlet fission effect highly improbable for the contact tetracene layer. Additionally, the presence of the silicon substrate results in a modification of the singlet-triplet gap by 144 meV. This change is solely attributed to the hybridization with excitons in silicon, which influences the exchange energy. Our results show that the dynamic dielectric screening caused by the substrate does not impact the singlet-triplet gap but alters the exciton binding energies.
\end{abstract}

\maketitle

\section{Introduction}

Singlet fission (SF) in organic crystals, first observed in 1965 \cite{doi:10.1063/1.1695695}, has garnered increasing attention in recent years due to its potential applications in photovoltaic devices. The main feature of this effect is the down-conversion of one singlet excited state into two long-lived triplet excitons avoiding thermal losses. These two triplet excitons can then be converted into four charge carriers, thus increasing the yield of charge carriers per photon \cite{Rao2017}. Several photovoltaic devices based on the SF effect have been proposed. One of the most promising proposals considers SF in hybrid inorganic-organic semiconductor (HIOS) heterostructures, such as tetracene-silicon interfaces, which we also study in this work. \cite{MacQueen, Einzinger2019}. The chain of transitions that leads to the generation of electron-hole pair in silicon as a result of SF in tetracene can be schematically written as:

\begin{equation}
S_0S_0 \xrightarrow{h\nu} S_0S_1 \xrightarrow{k_{fis}} T_1T_1 \xrightarrow{k_{tr}} S_0 + 2h_{Si}^+ + 2e_{Si}^-    
\label{chain}
\end{equation}
Here, a photon with energy $h\nu$ excites the tetracene molecule from its ground state $S_0$ to the singlet state $S_1$. Through the SF effect, the latter decays into two triplets $T_1$ with the rate $k_{fis}$. The energy transfer of the triplet states across the interface, with the rate constant $k_{tr}$, generates electron-hole pairs in silicon. In this setup, the organic layer enables efficient generation of excitons through SF, while the inorganic layer provides efficient separation and transport of charge carriers to the electrodes. The energy of the triplet exciton in the organic semiconductor should exceed the band gap of the inorganic semiconductor to facilitate resonant energy transfer \cite{Rao2017}. This is a critical requirement for attaining high quantum efficiency and surpassing the limitations imposed by the Shockley-Queisser limit \cite{Shockley-Queisser}. The triplet excitons should also be located in close proximity to the silicon substrate due to the short-range nature of the Dexter exciton transfer. In what follows, we refer to the tetracene molecular layer in contact with silicon as the contact layer. The chain of the reactions \eqref{chain} indicates that the concentration of electron-hole pairs in silicon is also dependent on the SF rate $k_{fis}$ determining the concentration of triplets in the contact layer.  Note that the intralayer diffusion of singlet excitons greatly exceeds the interlayer one \cite{Akselrod2014}. As a result, the majority of singlet excitons, once generated, are unlikely to escape the layer in which they were generated within the timeframe of their lifetimes.  This motivates interest in studying excitons in an organic semiconductor that are specifically located in the contact layer.

The aim of this work is to estimate the effect of the silicon substrate on the exciton binding energies and its implications for the SF effect in the tetracene contact layer. The thermodynamics of SF reads that the rate of SF is larger when the singlet energy is slightly larger (exothermic SF), or at least not much smaller (endothermic SF) than twice the triplet exciton energy \cite{Rao2017, Smith2010}. This states so-called thermodynamic and kinetic conditions for SF \cite{Casanova2018}. Another requirement concerns the wave function of excitons: the wave function for $S_1$ should manifest a charge transfer character, ensuring a significant overlap with the wave functions of the intermediate states along the pathway to the eventual triplet states \cite{doi:10.1146/annurev-physchem-040214-121235, PhysRevLett.110.226402, doi:10.1063/1.4892793, Zeng2014}. Note that an excessively long-range charge transfer character can be detrimental to SF. This is because such states are weakly coupled to the multiexciton triplet-pair manifold, as has been previously established in Ref. \citenum{doi:10.1063/1.4892793}. When the substrate induces alterations in the energies and wave functions of singlet excitons via the dynamic dielectric screening and orbital hybridization, it consequently exerts an influence on the SF process.

The exciton binding energy is determined by both the strength of Coulomb coupling between electrons and holes and their dispersion law (band structure, effective masses etc.). Modifying dielectric properties of the media by surfaces, interfaces, and nanostructuring allows for engineering excitonic features. The effect of the dielectric screening on the exciton binding energy has been thoroughly studied in the context of inorganic semiconductor materials in 3D, 2D, 1D and 0D III-V semiconductors \cite{haug2009quantum, klingshirn2007semiconductor} as well as in recently developed 2D van-der-Waals heterostructures \cite{PhysRevLett.113.076802, PhysRevB.97.035425, Andersen2015}, but less so for the HIOS heterostructures. Note, that the Wannier-Mott excitons in inorganic semiconductors and Frenkel excitons in organic semiconductors exhibit several distinctive features. For instance, the local-field effects play an exceptional role only in the case of Frenkel excitons \cite{Onida, PhysRevB.103.115203}.

 In this work, we study the impact of the substrate on the exciton binding energies and SF effect within the contact tetracene layer, elucidating specifically  the role of the dynamic dielectric screening and orbital hybridization between excitons in organic and inorganic semiconductors. Using a combination of the GW theory and Bethe-Salpeter equation (BSE) \cite{PhysRevB.62.4927, Onida, BGW, doi:10.1063/1.4983126}, we perform a series of computational experiments with several models representing the effect of substrate. One relies on a brute-force approach that utilizes slab models and large supercells that encompass both the tetracene and silicon slabs. Another is based on the so-called "Add-chi" method \cite{Xuan2019}, which can be potentially used with the dielectric embedding \cite{Liu2019, Liu2020} to reduce the sizes of supercells and, consequently, computational burden. The "Add-chi" and dielectric embedding techniques rely on the additivity of the polarizability matrix, which holds true under specific conditions. They have previously been employed in GW computations, effectively reducing computational expenses. These methods are particularly well-suited for heterogeneous structures characterized by weak van der Waals interactions between components, where orbital hybridization can be neglected. Notably, this study represents marks the first instance of combining dielectric embedding with BSE. The combination of GW and BSE methods has previously demonstrated success in predicting the excitonic properties of various materials. For instance, it has been employed to investigate the influence of crystal packing on exciton probability distributions \cite{doi:10.1063/1.5027553} or excitonic signatures in optical absorption spectra of inorganic semiconductors \cite{Wing}.

\section{Results and discussion}

\subsection{Atomic model of tetracene-silicon interface}

\begin{figure}[t]
    \subfigure[]{\includegraphics[width=0.9\linewidth]{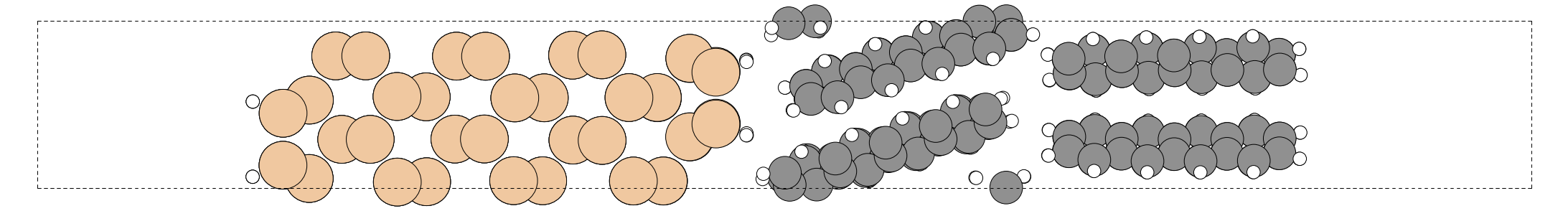}}
    \subfigure[]{\includegraphics[width=0.75\linewidth]{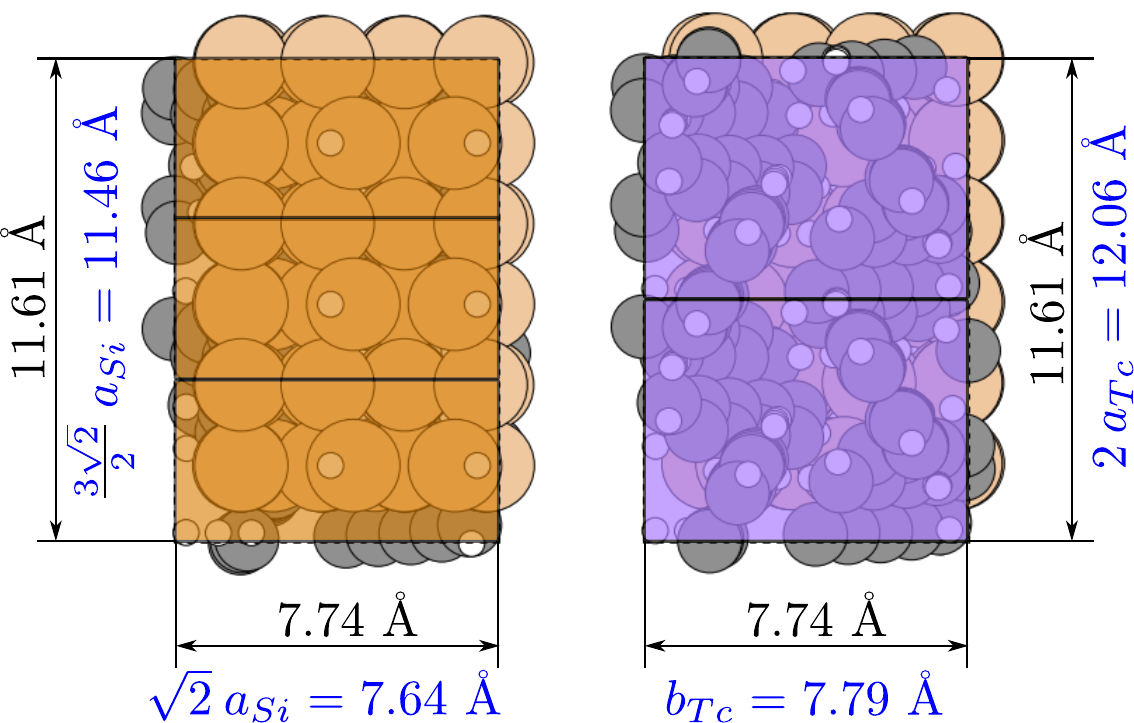}}
    \caption{a) Atomic structure of the van-der-Waals interface between the crystalline silicon with the 1x2 reconstructed (100) surface and tetracene after atomic relaxation. b) Alignment of the silicon and tetracene unit cells at the interface. Numbers in blue color denote sizes of unit cells for bulk materials published in the literature, black color numbers correspond to the sizes of the super-cell after relaxation computed in this work. $a_{Si}$, $a_{Tc}$, and $b_{Tc}$ are crystal lattices of bulk silicon and tetracene respectively.}
    \label{fig:slab}
\end{figure}

In this work, we consider tetracene deposited onto neutral hydrogen-passivated 1x2 reconstructed (100) silicon surface in the \enquote{upright-standing} configuration \cite{Betti2007, Niederhausen2020}. This configuration has been previously grown and characterized experimentally \cite{PhysRevB.74.205326}. The 1x2 reconstructed Si(100) surface has a rectangular unit cell with lateral sizes of $\sqrt{2} a_{Si}$ and $(\sqrt{2} / 2)  a_{Si} $ along the axes $[110]$ and $[\bar{1}10]$ respectively, where $a_{Si}$ is the lattice constant of bulk silicon. The minimal unit cell spans giving the best match of sublattices are $1 \times 2$ for tetracene and $1 \times 3$ for Si, shown schematically in Fig. \ref{fig:slab}. All computations in this work are performed for a slab model containing 16 atomic layers of silicon and two molecular layers of tetracene. We need at least two molecular layers to simulate accurately the dielectric environment for the tetracene molecules at the interface.

The atomic coordinates have been obtained from the geometry optimization within DFT with a plane-wave basis set and norm-conservative pseudo-potentials \cite{Hamann, PseudoDojo}. The computations were performed using Quantum Espresso, the plane-wave DFT software \cite{Giannozzi_2017, QE}. The dispersion forces, responsible for the physisorption of tetracene on the Si surface, are introduced in the model via the non-local exchange-correlation functional vdW-DF2-C09 \cite{C09}. The computations have been performed on the 4x3x1 Monkhorst-Pack k-space grid, using a kinetic energy cutoff of 80 Ry for wavefunctions and of 320 Ry for charge densities. This choice is justified by a series of convergence tests (see Supplementary info). For the geometry optimization, we used the use Broyden-Fletcher-Goldfarb-Shanno quasi-newton algorithm with variable cell parameters. The system is periodic only in two in-plane dimensions. The effect of the periodic boundary conditions in the third dimension was canceled by the dipole correction \cite{PhysRevB.59.12301}.

The band gap and exciton binding energy are both properties of the excited state. While the static DFT is fundamentally a ground state theory, it can still provide rough estimates of these properties for specific material systems by employing well-designed approximated exchange-correlation functionals, such as range-separated hybrid functionals \cite{Yang2023}. More accurate predictions in a more systematic way, however, can be achieved by employing specialized excited state theories like time-dependent density-functional theory or post-Hartree-Fock methods, such as multi-configurational self-consistent field \cite{ROOS1980157}, coupled clusters method \cite{CC}, configuration interaction methods \cite{szabo}, or many-body Green's function approach with GW approximation \cite{Onida}. The latter works exceptionally well for large extended systems when combined with techniques like "Add-chi" \cite{Xuan2019} or dielectric-embedding methods \cite{Liu2019, Liu2020}. Within this method, the quasiparticle energies are determined by the poles of the retarded Green's functions, which, in turn, are solutions of the Dyson equation. In the general case, this equation can be solved self-consistently as a part of the closed system of Hedin's equations \cite{Louie, Onida}. Detailed information on the application of the GW method to the tetracene-silicon interface can be found in Ref. \citenum{Klymenko_2022}. 

\begin{figure}[t]
    \centering
    \includegraphics[width=0.7\linewidth]{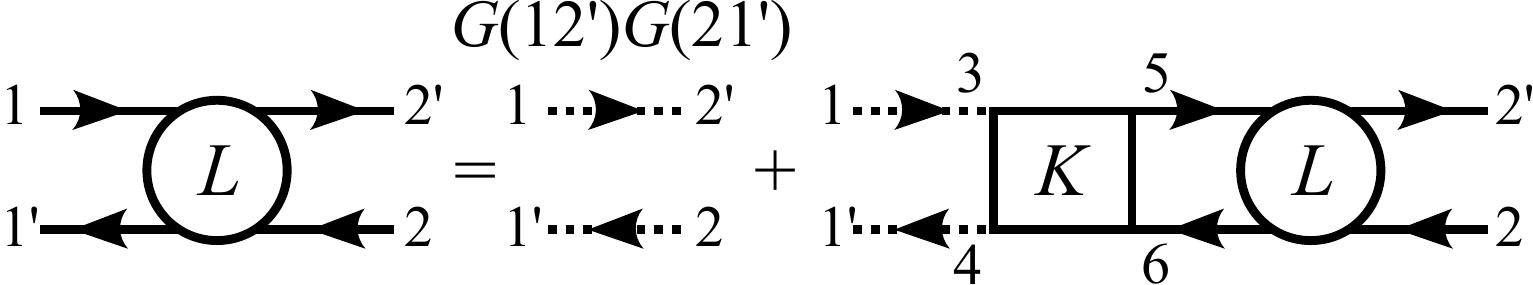}
    \caption{The diagrammatic representation of the Bethe-Salpeter equation.}
    \label{fig:feynman_diag}
\end{figure}

After obtaining the quasiparticle energies and orbitals through the GW method, the exciton binding energies and exciton wave functions can be determined in the subsequent iteration of Hedin's equation by solving BSE, which, in the general case, reads \cite{Strinati1988, PhysRevB.62.4927}:

\begin{equation}
    \begin{split}
        L\left(12;1'2'\right) = &G\left(12'\right) G\left(21'\right) + \int d\left( 3456\right) \times\\
        &G\left(13 \right)G\left(41' \right) K\left(35;46 \right)  L\left(62;52'\right)
    \end{split}
    \label{BSE}
\end{equation}
where each number in the parenthesis denotes the
set of variables consisting of position, spin, and time coordinates, $(1)=(\mathbf{r}_1 , s_1, t_1)$, $G\left(12'\right)$ is the non-interacting one-particle propagator (computed, for instance, by the GW method),  $L\left(12;1'2'\right)$ is the electron-hole correlation function and $K\left(35;46 \right)$ is the electron-hole interaction kernel. The diagrammatic representation of Eq. \eqref{BSE} is shown in Fig. \ref{fig:feynman_diag}. The poles of the functions $L\left(12;1'2'\right)$ correspond to the exciton energy levels. The kernel consists of the exchange and screened direct Coulomb interaction \cite{PhysRevB.62.4927}:

\begin{equation}
    \begin{split}
        K\left(35;46 \right) = & \delta\left(34\right) \delta \left(56\right) v \left(36\right) - \\
        &\delta\left(36\right) \delta \left(45\right) W \left(34\right) 
    \end{split}
    \label{kernel}
\end{equation}
where  $v\left(36\right)$ and $W\left(34\right)$ are bare and screened Coulomb potentials respectively. 

The integral equation \eqref{BSE} can be converted into a linear algebra problem by expressing all its components in matrix form and transitioning from the time domain to the frequency domain through a Fourier transform. The matrix representations were derived using the orbitals obtained from the GW method as the basis set. In the matrix form, the screened Coulomb potential is given by:

\begin{equation}
    W = \epsilon^{-1} v = \left[1-v \chi \right]^{-1} v
    \label{hedin2}
\end{equation}
where: $\epsilon^{-1}$ is the dielectric matrix inverse, $\chi$ is the irreducible polarizability of the medium, and $v$ is the matrix representation of the bare Coulomb potential. The explicit expressions for the matrix representations of $\chi$ and $v$, utilizing the GW orbitals or Kohn-Sham orbitals as the basis set, can be found in Refs. \citenum{Onida} and \citenum{PhysRevB.62.4927}.

In this work, we, for the first time, apply the dielectric embedding technique with BSE to obtain estimates for the effect of dielectric screening, caused by the substrate, on excitons in tetracene. Let us consider the two-particle correlation function $L\left(1,2;1',2'\right)$ for a heterostructure formed by two slabs $A$ and $B$. In the general case, the initial and final positional coordinates can be located in any of two materials:  $\{1,2,1',2'\} \in A \cup B$. If our focus is solely on excitons in one of the materials, say $A$, and the hybridization between Kohn-Sham orbitals in $A$ and $B$ is negligibly small, then the set of coordinates is $\{1,2,1',2'\} \in A$. In other words, the electron-hole pair before and after interaction remains in the material $A$. The only possible effect that material B can cause on the excitons in A is the contribution to the dielectric screening of the Coulomb potential. The dielectric embedding is based on the assumption that the polarizability matrix $\chi$ is additive and can be computed independently for isolated slabs constituting the heterostructure \cite{Liu2019, Liu2020}:
\begin{equation}
\chi \approx \chi^{A} + \chi^{B}.
\label{chi}
\end{equation}

To save computational resources, the contributions $\chi^{A}$ and $\chi^{B}$ can be computed by utilizing supercells with sizes adjusted to geometrical parameters of the isolated slabs $A$ and $B$. In this case, before summing them in the reciprocal space, it is essential to adjust these contributions by transforming them into real space, incorporating zero-padding in real space, and then reverting them back to reciprocal space. This procedure forms the essence of the dielectric embedding method, which has been previously employed with the GW method \cite{Liu2019, Liu2020}.

In the subsequent discussion, we will denote the solutions of the Bethe-Salpeter equation obtained without the dielectric embedding method as the "double-slab model" (DSM), while the alternative case will be referred to as the "single-slab model" (SSM).

Since we are only interested in the poles of the functions $L\left(12;1'2'\right)$, BSE in the Tamm-Dancoff approximation reduces to the following eigenvalue problem: \cite{PhysRevB.62.4927}:

\begin{equation}
    \left(E_{i}^{QP} - E_{j}^{QP} \right) A_{jiji} + \sum_{j'i'} K_{j'i'ji} A_{j'i'ji} = E_{ji} A_{jiji}
    \label{eigenvalues}
\end{equation}
where $E_{i}^{QP} $ are quasiparticle energies obtained from GW computations, $A_{j'i'ji}$ represents the eigenvectors, which determine the amplitudes in the representation of the two-particle wave function, and $E_{ji}$ is the eigenvalues that can be interpreted as exciton energies, measured relative to the fundamental band gap, $K_{j'i'ji}$ is the matrix representation of the kernel defined by Eq. \eqref{kernel} in the basis set of the GW orbitals, and indices $i$ and $j$ run over conduction and valance bands respectively. The matrix $A_{ji}$ allows for recovering the electron-hole correlation function:

\begin{equation}
     \Psi_{ij}\left(\mathbf{r}_h, \mathbf{r}_e \right)= \sum_{i'j'} A_{j'i'ji} \psi_{i'}\left(\mathbf{r}_e \right) \psi_{j'}\left(\mathbf{r}_h \right)
    \label{exciton_wf}
\end{equation}
where $\psi_{i'}\left(\mathbf{r}_e \right)$ and $\psi_{j'}\left(\mathbf{r}_h \right)$ are GW wave functions for electrons and holes respectively.

All the computations related to GW and BSE have been performed using BerkeleyGW \cite{BGW}. BSE has been solved with the Tamm-Dancoff approximation. In this work we use two techniques to reduce the number of the bands explicitly participating in the computations: first, we apply the modified static remainder approach \cite{Deslippe} and use the extrapolation technique based on fitting the Coulomb-hole self-energy by a hyperbolic function \cite{Tamblyn}. After a series of convergence tests discussed in Supplementary information we have derived the following parameters: the kinetic energy cutoff for the dielectric matrix is of 15 Ry, 534 unoccupied orbitals (1200 orbitals in total) were used to build matrix representation of the Green's functions, and the k-grid is same as for the DFT calculations.

\begin{figure}[t]
    \centering
    \includegraphics[width=0.7\linewidth]{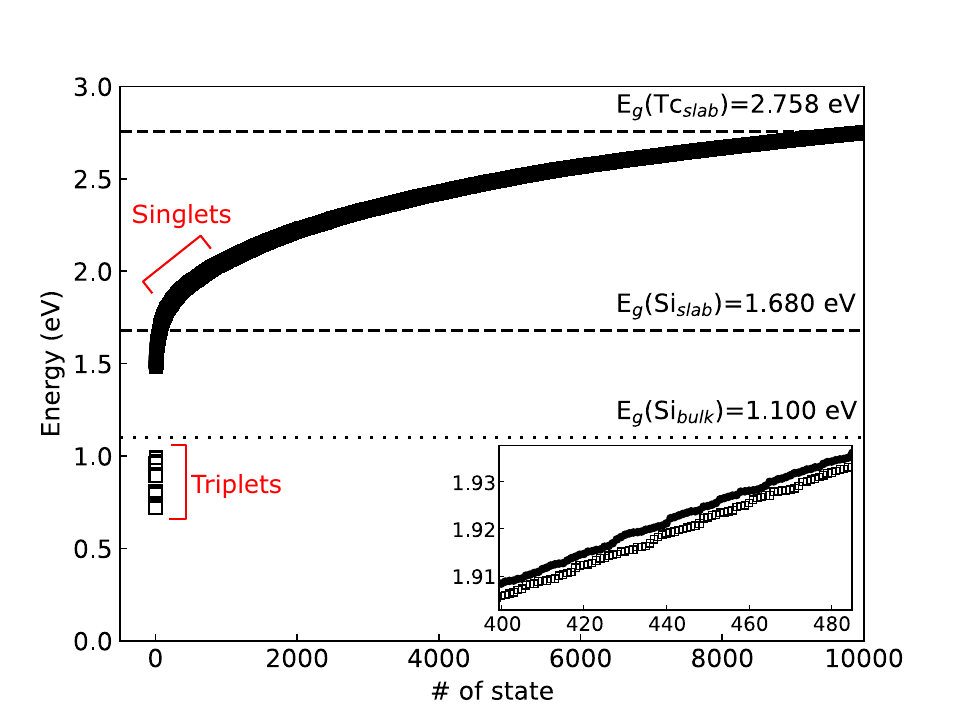}
    \caption{Energy levels of singlet ($\sbullet[1.5]$) and triplet ($\square$) excitations computed by solving numerically Bethe-Salpeter equation for the supercell containing contacting silicon and tetracene slabs.  The inset panel provides a close-up view of a segment of the spectrum, highlighting the region where the tetracene singlet excitons are located. The figure also contains the values of band gaps, $E_g$, of the considered materials.}
    \label{fig:eigenvalues}
\end{figure}

\subsection{Exciton binding energies}

Solving the eigenvalue problem given by Eq. \eqref{eigenvalues} yields a spectrum of exciton binding energies, shown in Fig. \ref{fig:eigenvalues} for DSM. The resulting spectrum consists of about ten thousand eigenvalues within the energy interval spanning the band gap of tetracene and comprises both discrete and quasi-continuous parts (numerical solution of Eq. \eqref{eigenvalues} is possible for a finite number of wave vectors that leads to the discretization of the continuous part of the spectrum). The discrete and continuous parts correspond to localized and delocalized exciton states respectively. The latter is related to correlated motion of unbound electrons and holes, which, in particular, manifests itself as the Sommerfeld enhancement in the absorption spectrum \cite{PhysRevB.38.3342}.  

For this study, our attention is directed toward excitons situated within the tetracene contacting layer. These particular exciton states have been identified in the spectrum through a projection of their corresponding exciton wavefunctions onto a bounding box that encapsulates this layer. Another discriminant that can be employed to distinguish excitons in tetracene from other types of excitons is their dipole moment. Results of our numerical experiments show that excitons in organic semiconductors typically exhibit dipole moments at least one order of magnitude larger than Wannier-Mott excitons.

Our results obtained from DSM show that, while the triplet excitons in tetracene are located in the fundamental band gap, the singlet exciton states in tetracene are embedded into the continuous spectrum of the unbound exciton states in silicon (see Fig. \ref{fig:eigenvalues}). The singlet exciton energy levels are broadened as a result of interaction with the silicon substrate, whereas the triplet states remain very narrow. To avoid any ambiguity associated with this broadening, henceforth, when referencing the energy of a singlet exciton, we specifically imply the energy of the singlet exciton that is characterized by the maximal overlap with the tetracene contacting layer. 

\begin{figure}[]
    \centering
    \subfigure[]{\includegraphics[height=0.65\linewidth]{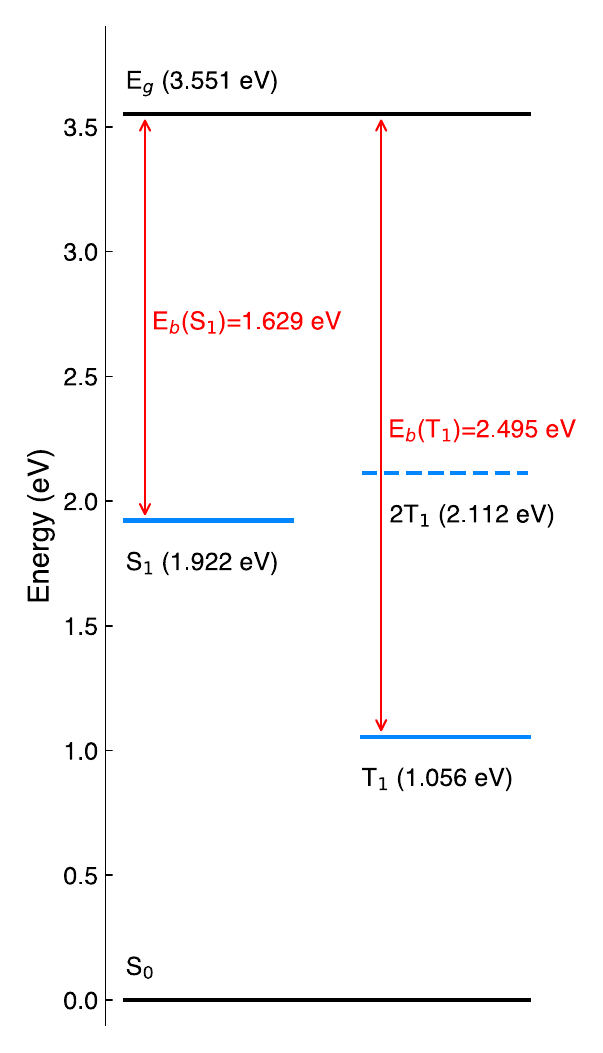}}
    \subfigure[]{\includegraphics[height=0.65\linewidth]{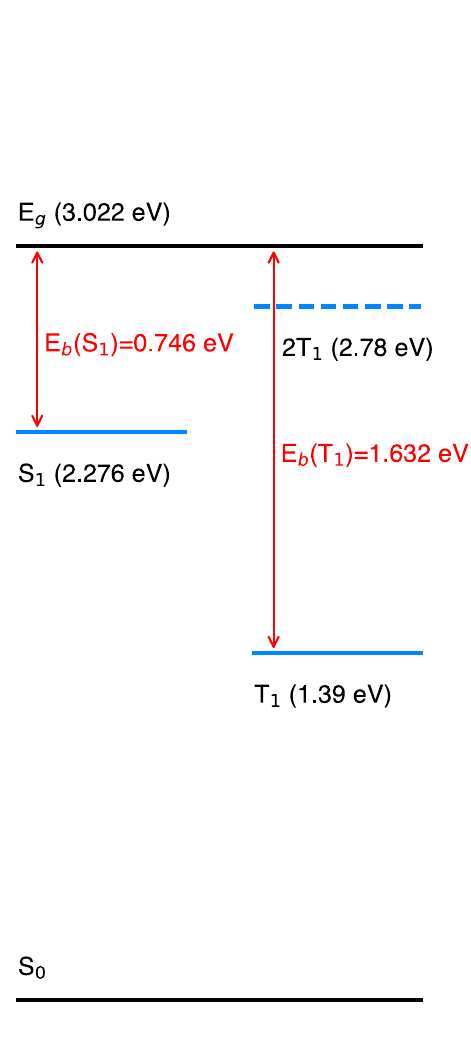}}
    \subfigure[]{\includegraphics[height=0.65\linewidth]{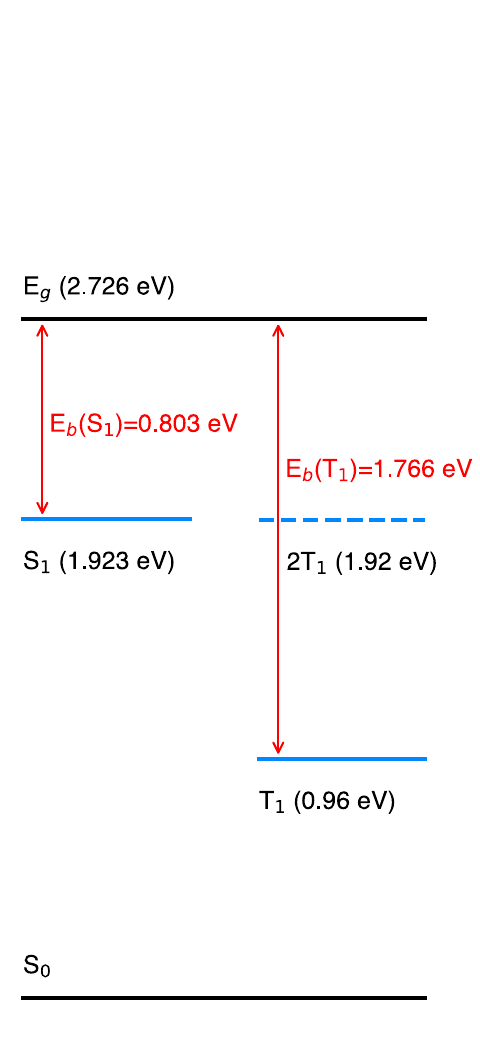}}
    \caption{Exciton energy levels (blue lines) and band gaps for the contact tetracene layer computed for (a) isolated tetracene slab and (b) tetracene-silicon interface within SSM and (c) tetracene-silicon interface within DSM. The binding energies for the exciton singlet state $S_1$ and triplet state $T_1$ are marked by red arrows with an annotating text.}
    \label{fig:levels}
\end{figure}

In Fig. \ref{fig:levels}, we compare the exciton binding energies, as indicated by the red arrows, for three cases: an isolated tetracene slab, a tetracene-silicon heterostructure treated with SSM, and a tetracene-silicon heterostructure treated with DSM. Comparing the isolated slab with SSM shows solely the effect of dielectric screening, while DSM takes into account both dielectric screening and orbital hybridization between slabs. 

\begin{figure}[t]
    \includegraphics[width=0.5\linewidth]{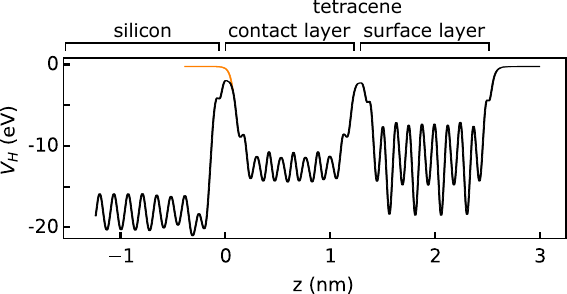}
    \caption{Potential barrier lowering caused by silicon substrate in the confinement potential of tetracene molecules in the contact layer. The orange line indicates the potential barrier for the case of the isolated slab.}
    \label{fig:pot_barrier}
\end{figure}

The results of the computations show that the presence of the silicon substrate decreases the exciton binding energy for both singlet and triplet states. This outcome is expected; we observe the same behavior in 2D inorganic semiconductors, where the exciton binding energy similarly decreases in the presence of a substrate due to the dielectric screening of the Coulomb coupling between electrons and holes \cite{PhysRevLett.113.076802, PhysRevB.97.035425, Andersen2015}. The reduction in binding energies is qualitatively captured by both SSM and DSM. However, quantitatively, these two models exhibit a slight discrepancy, especially in the case of the triplet: the triplet binding energy is 201 meV larger in the latter case, whereas the binding energy of the singlet exciton is only 57 meV larger for DSM. As a result, SSM underestimates the singlet-triplet energy gap by 144 meV. The singlet-triplet gap is determined by the exchange interaction energy, which is non-zero for triplet states and zero for singlet states. This exchange interaction is inherently short-range and depends on the overlap of the singly-occupied orbital. Consequently, it remains unaffected by long-range dynamical dielectric screening, as evident in the comparison between the results for the isolated slab and those for the SSM case in Figure \ref{fig:levels}. However, the singlet-triplet gap changes when accounting for exciton hybridization between silicon and tetracene, as observed in the comparison between the SSM and DSM cases in Figure \ref{fig:levels}.

Figure \ref{fig:levels} contains values of band gaps (depicted by the black lines) resulting from the GW calculations, which have been previously reported and analyzed in Ref. \citenum{Klymenko_2022}. The bandgap decreases when tetracene comes into contact with silicon due to the dynamic dielectric screening effect well reproduced by SSM. However, DSM predicts even a narrower bandgap as compared to SSM, and this can be attributed to the reduction of the confinement potential of the tetracene molecules in the contact layer. By subtracting the binding energies from the band gap energy of tetracene, we can obtain an estimate of the energy of the lowest singlet and triplet excited states relative to the ground state. In comparison to the isolated slab case, the energies of excitons increase slightly when the substrate is introduced in the SSM case. Conversely, in the DSM case, the singlet energy remains unchanged, while the triplet energy decreases by 60 meV. When comparing SSM and DSM, we can conclude that taking the hybridization of exciton states between tetracene and silicon into account leads to a reduction in exciton energies relative to the ground state for both singlets and triplets. This can be explained by the overall lowering of the potential barrier caused by silicon in the confinement potential of tetracene molecules, as illustrated in Fig. \ref{fig:pot_barrier}.

\begin{figure}[t]
    \centering
    \subfigure[]{\includegraphics[width=0.25\linewidth]{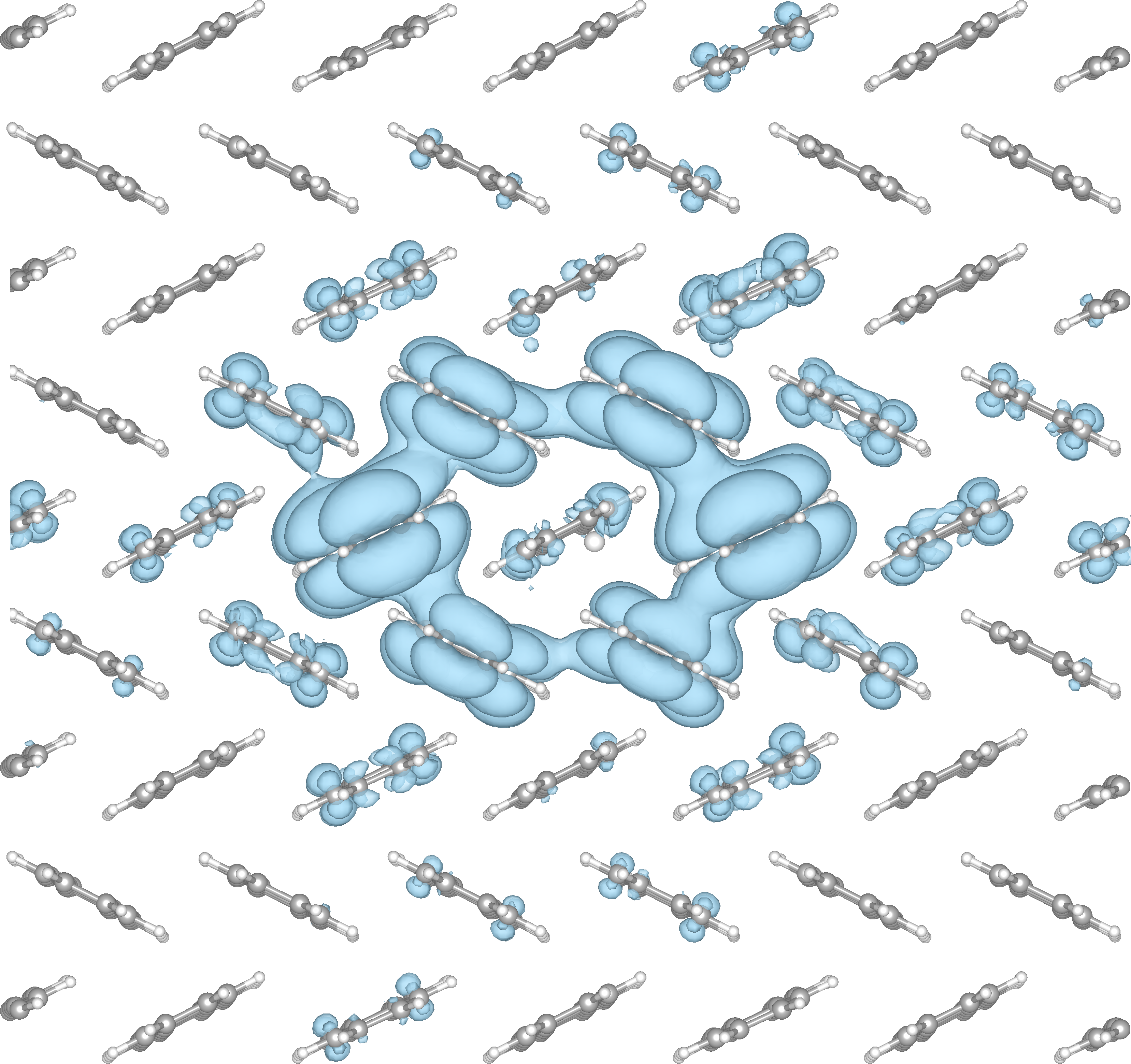}}\hspace{0.1cm}
    \subfigure[]{\includegraphics[width=0.25\linewidth]{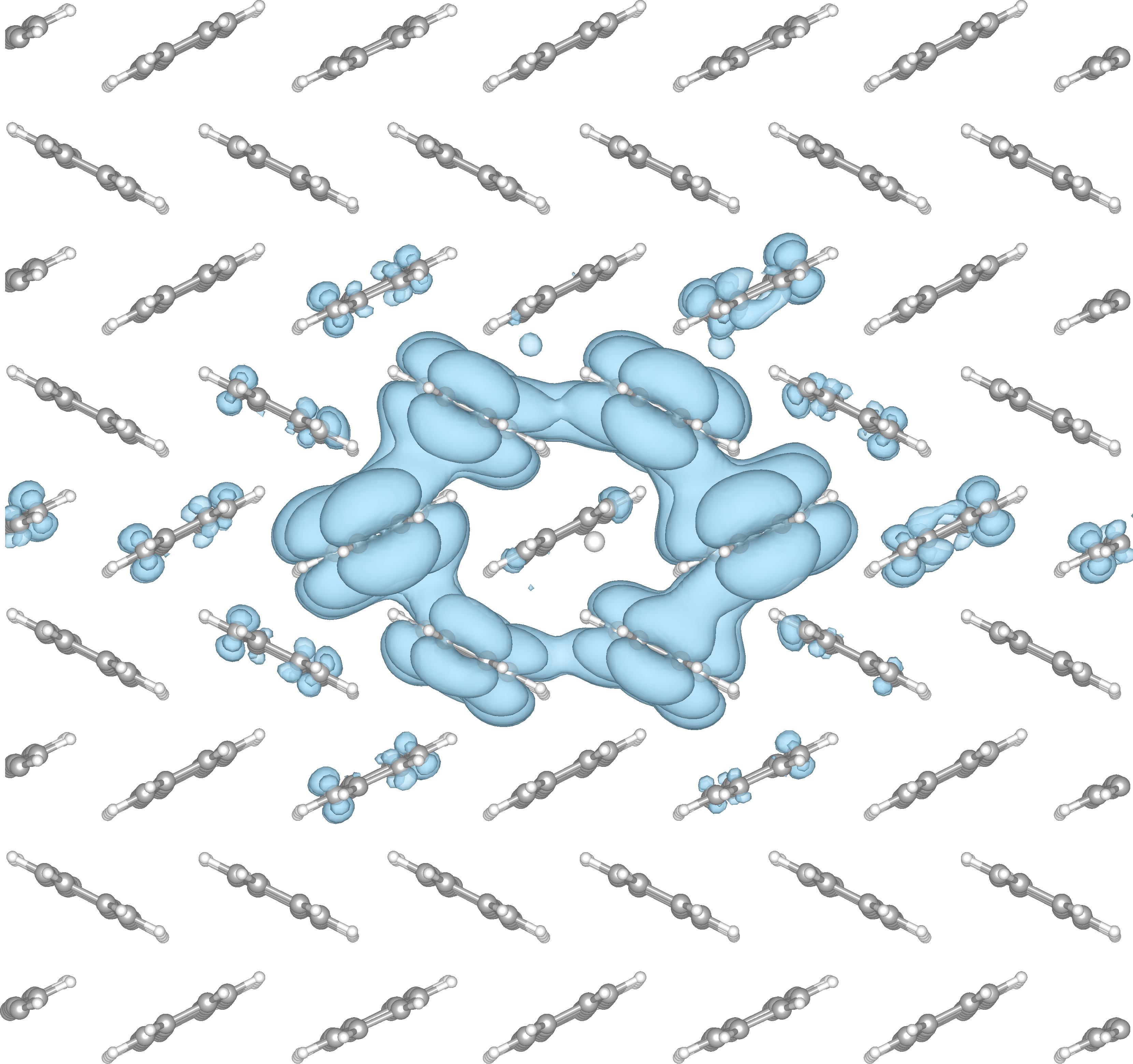}}\hspace{0.1cm}
    \subfigure[]{\includegraphics[width=0.25\linewidth]{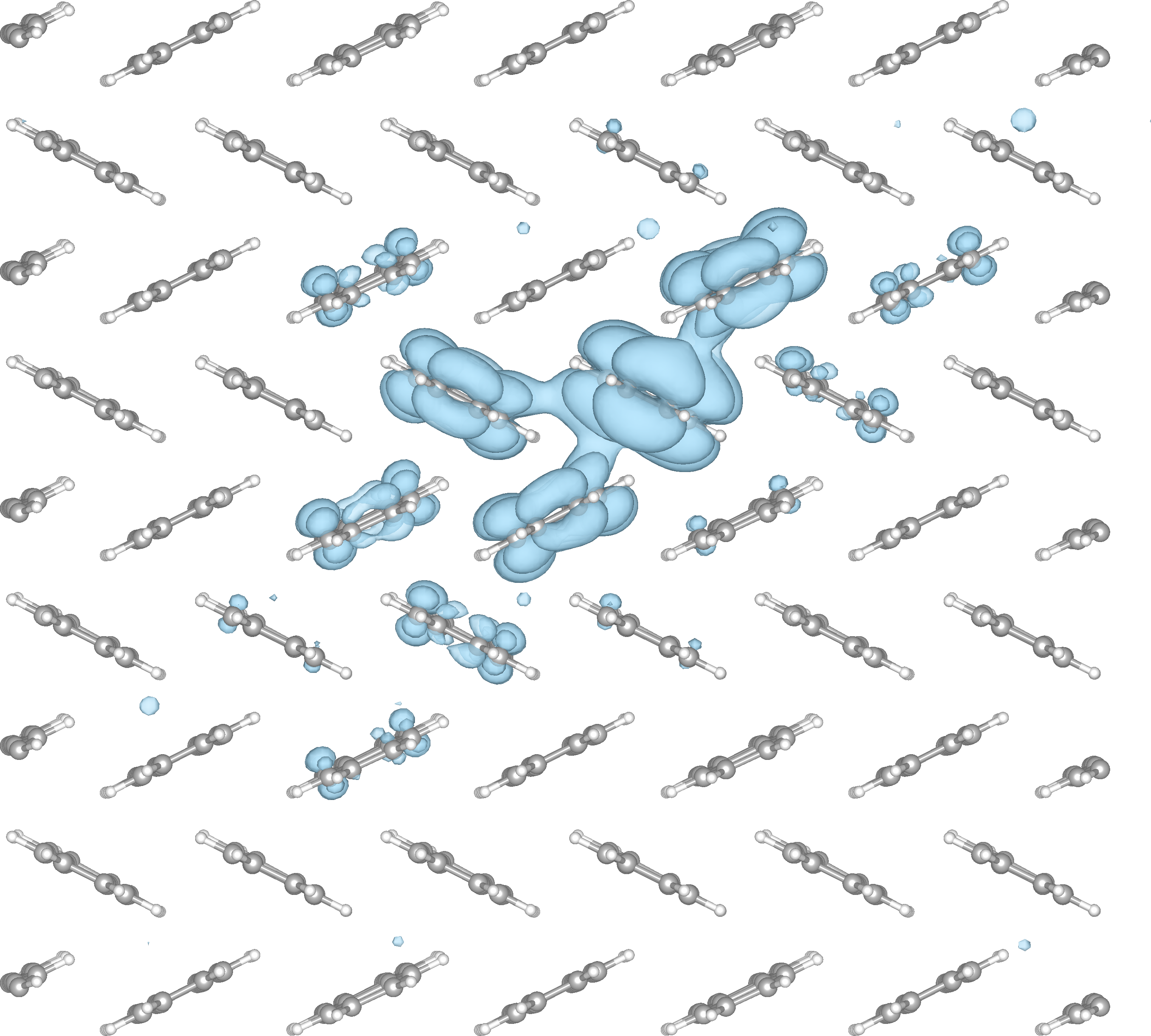}}
    
    \subfigure[]{\includegraphics[width=0.25\linewidth]{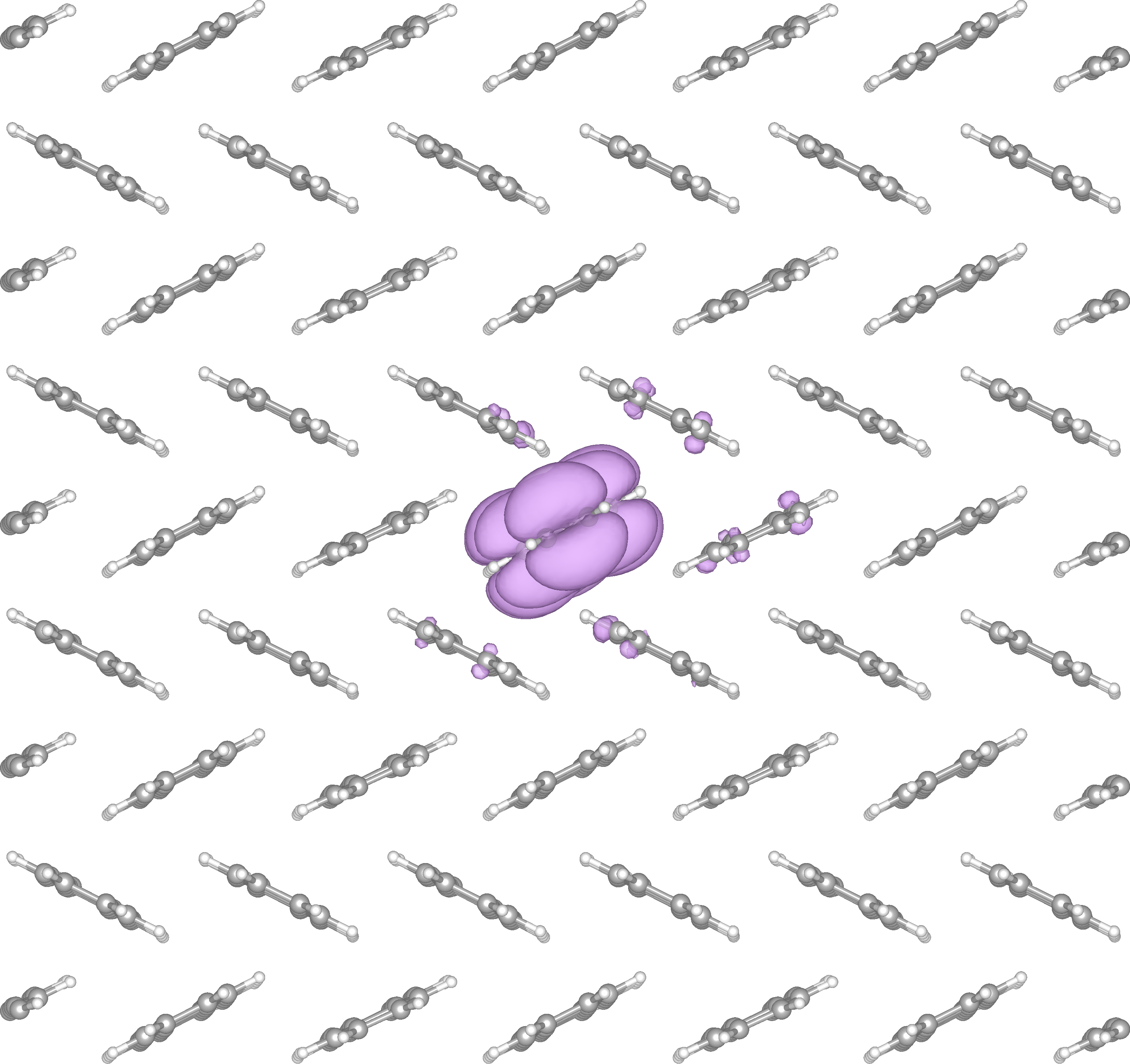}}\hspace{0.1cm}
    \subfigure[]{\includegraphics[width=0.25\linewidth]{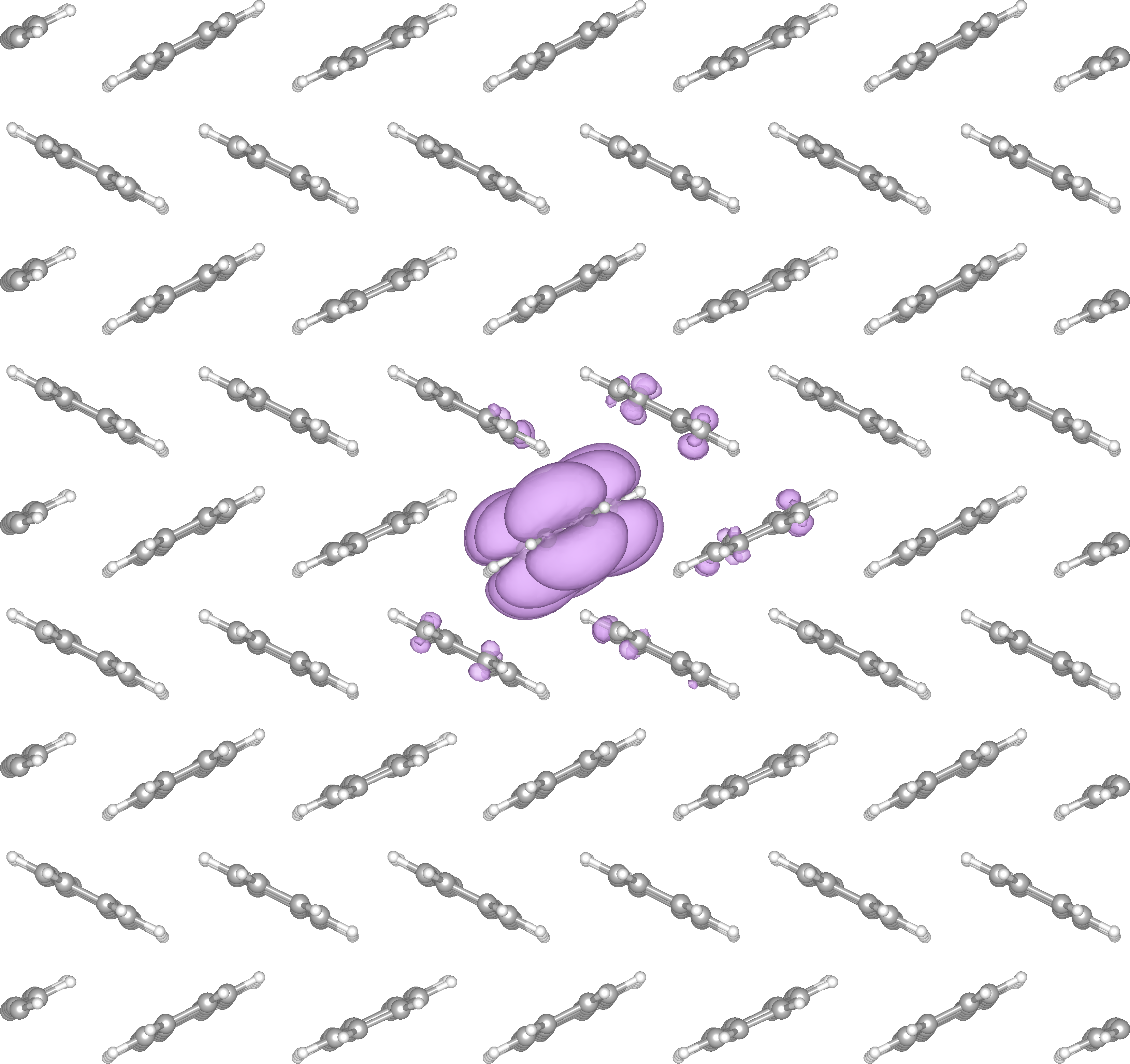}}\hspace{0.1cm}
    \subfigure[]{\includegraphics[width=0.25\linewidth]{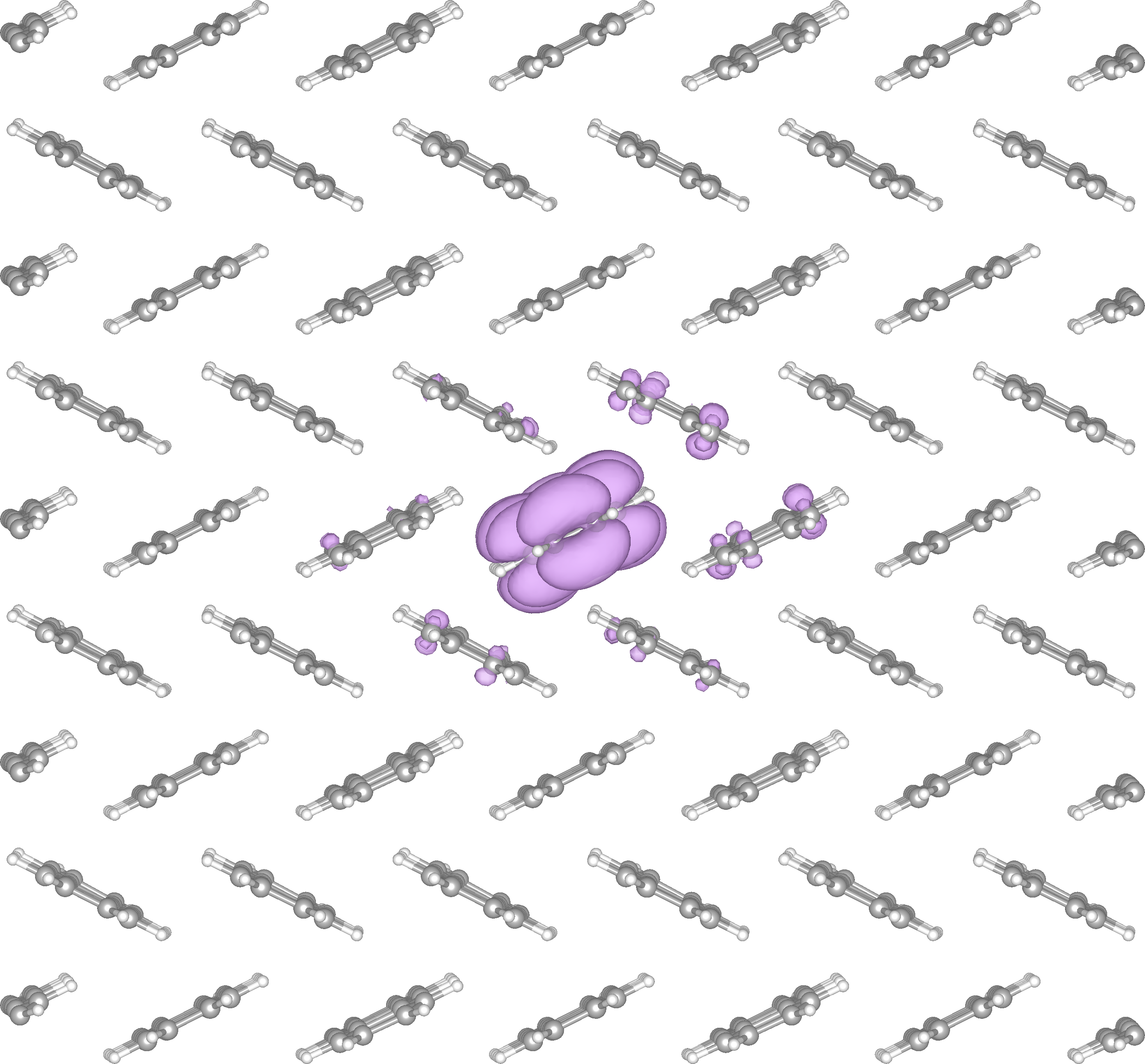}}
    \caption{Averaged electron density for the lowest singlet (a, b, c) and triplet (d, e, f) exciton states. The data on the panels (a) and (d) is for the isolated tetracene slab, (b) and (e) is for the tetracene on the silicon substrate computed with SSM, and (c) and (f) is for the tetracene on the silicon substrate computed with DSM. The isosurface of the electron density is plotted at a value of 0.005 of its maximum.}
    \label{fig:wf}
\end{figure}

The silicon substrate changes the bandgap more dramatically compared to the exciton energies. This is because uncorrelated electrons and holes are charged particles, while the bounded electron-hole pair is less sensitive to the external electrostatic environment due to mutual electrostatic screening.

For the contact tetracene layer, SSM gives the singlet exciton energy of 2.276 eV and bandgap of 3.022 eV being in good agreement with the results for the bulk tetracene published in Ref. \citenum{MacQueen} (S$_1$=2.2 eV and E$_g$=3.0 eV). It means the screening of the exciton produced by tetracene molecules is of the same order of magnitude as the screening by the silicon substrate. The discrepancy in the triplet energy is more pronounced in these cases (1.39 eV and 1.1 eV respectively). It can be associated with well-known inaccuracy of the GW-BSE method in describing the triplet spin states - it is known from the computations based on the configuration interaction method that the singlet state is well approximated by a single configuration, while the triplet state is a mixture of several configurations that can not be reproduced within the many-body perturbation theory approach.

\subsection{Exciton delocalization and charge transfer character}

To visualize spatial characteristics of excitons, we compute the electron probability density from the electron-hole correlation function $\Psi\left(\mathbf{r}_h, \mathbf{r}_e \right)$, see Eq. \eqref{exciton_wf}. Since our slab model is periodic in two dimensions but finite in the third dimension, to compute this function we fix the hole coordinates in the x-y plane, where periodic boundary conditions are applied, and then average over its positions along the confinement direction (along axis $z$):

\begin{equation}
    \left| \Psi\left(\mathbf{r}_e \right) \right|^2 = \frac{1}{L} \int d\mathbf{r}_h \left| \Psi\left(\mathbf{r}_h, \mathbf{r}_e \right) \right|^2 \delta\left(x - x_0 \right)\delta\left(y - y_0 \right), 
    \label{averaged}
\end{equation}
where $\mathbf{r}_h=(x,y,z)$ and $(x=x_0,\, y=y_0)$ defines the geometric locus of points forming the line that is perpendicular to the slabs and passes through the center of the supercell, and $L$ represents the extend of the supercell along the z-axis. With Eq. \eqref{averaged}, we compute the averaged electron probability density for excitons in the contact tetracene layer for the three cases discussed above - isolated slab, SSM, and DSM. The densities shown in Fig. \ref{fig:wf} are displayed from the viewpoint located at the plane of the interface, directed along the molecular axis, rather than being perpendicular to that plane. 

\begin{figure}[t]
    \centering
    \subfigure[]{\includegraphics[height=0.35\linewidth]{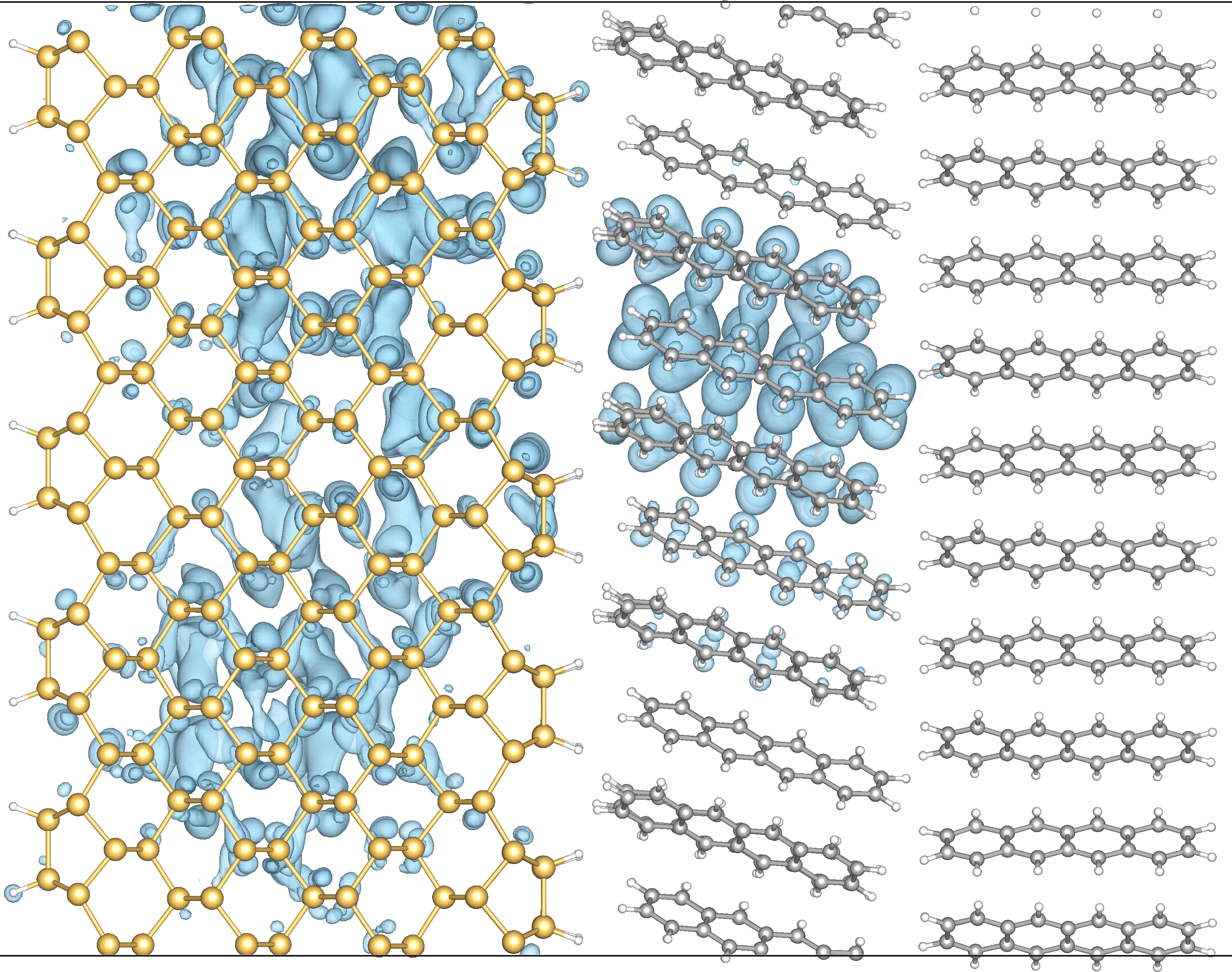}} \hspace{0.3cm}
    \subfigure[]{\includegraphics[height=0.35\linewidth]{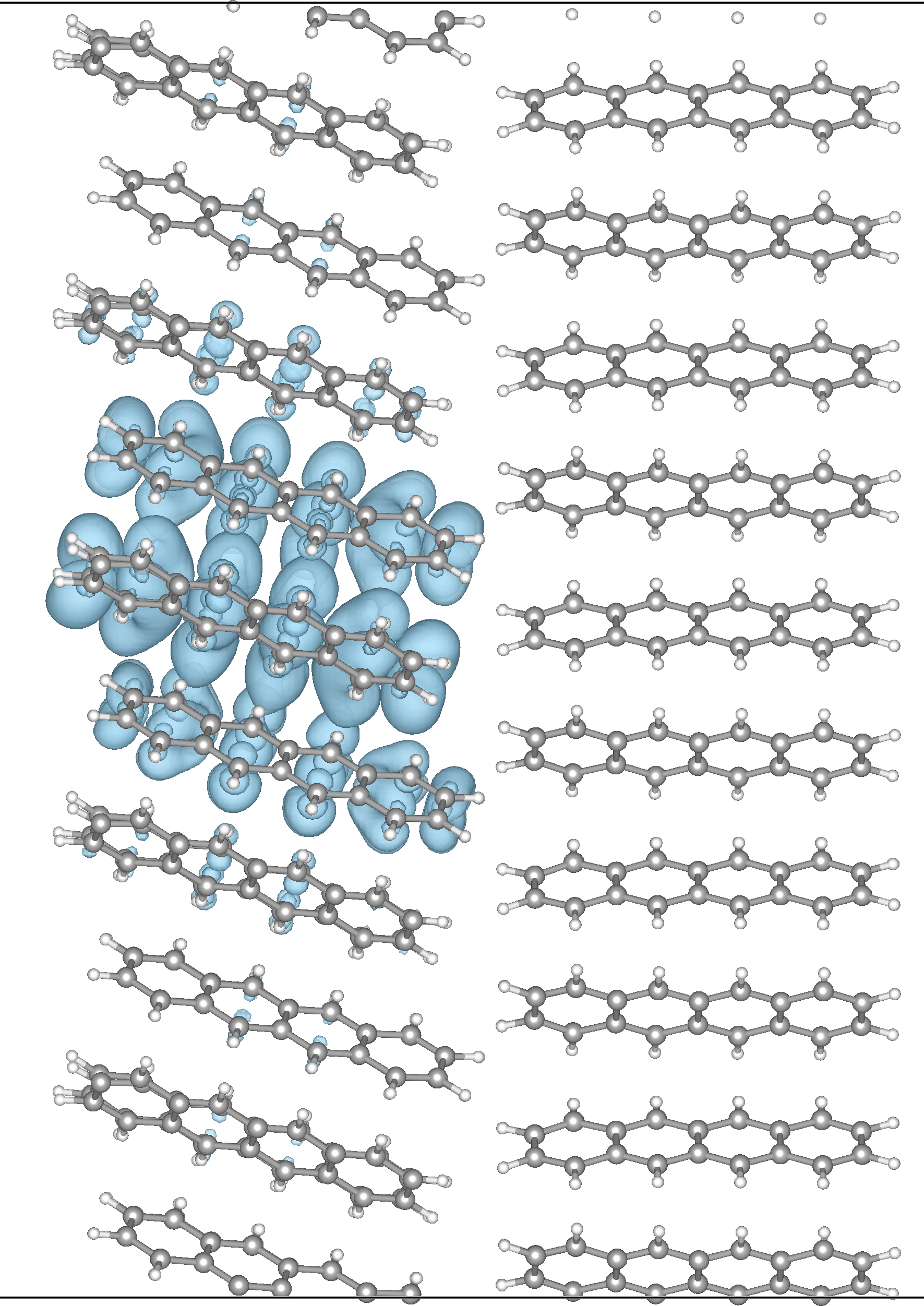}}
    \subfigure[]{\includegraphics[height=0.35\linewidth]{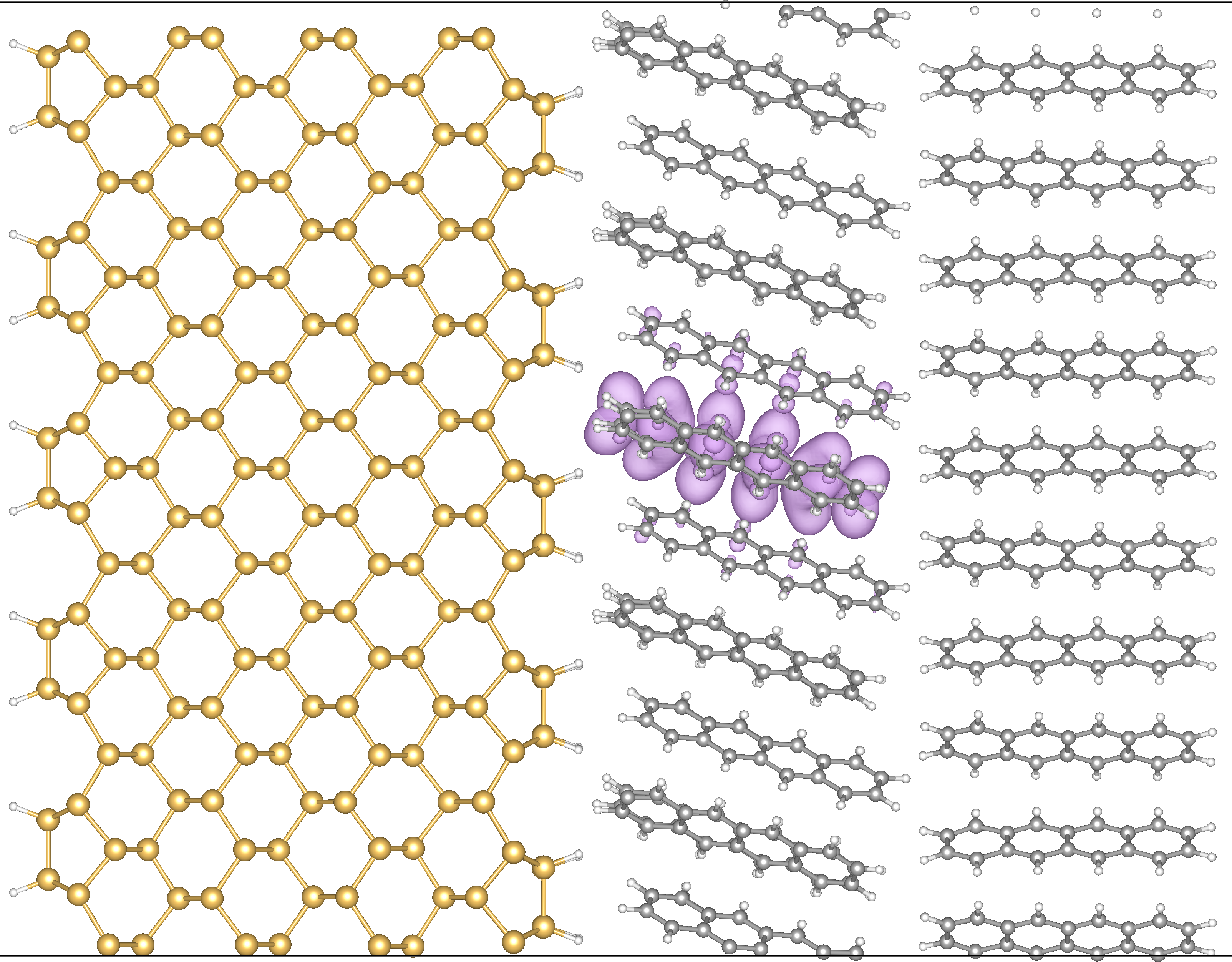}} \hspace{0.3cm}
    \subfigure[]{\includegraphics[height=0.35\linewidth]{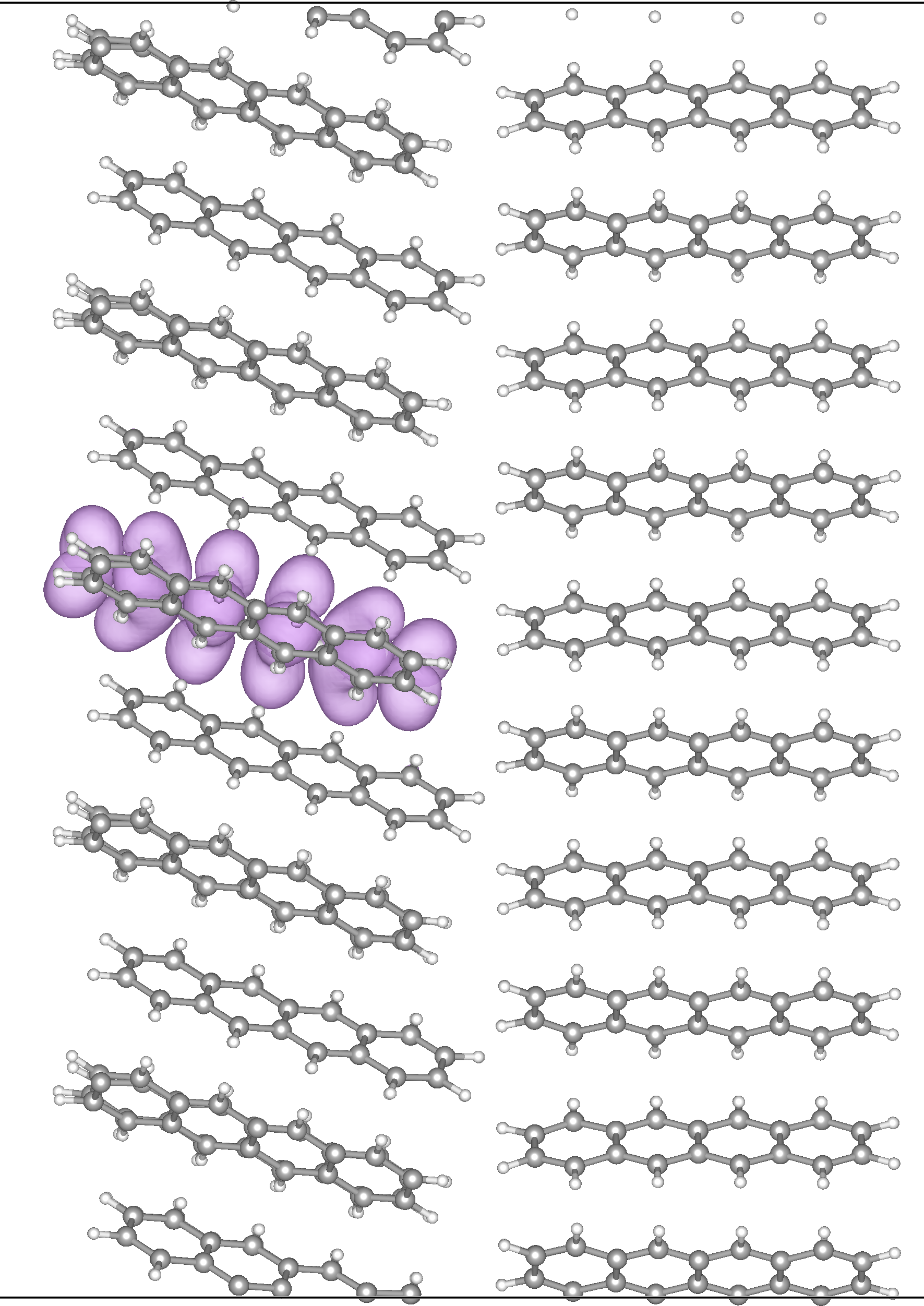}}
    \caption{Averaged electron density for the lowest singlet (a, b) and triplet (c, d) exciton states. The data on the panels (a) and (c) is for DSM, and the data on the panels (b) and (d) is for SSM. The isosurface of the electron density is plotted at a value of 0.01 of its maximum.}
    \label{fig:wf1}
\end{figure}

As is expected, the singlet exciton is more delocalized compared to the triplet exciton. The results for the isolated tetracene slab and SSM exhibit remarkable similarity. Both predict that for the singlet state, most of the electron density is located on the first nearest neighbor molecules, while for the triplet state, the electron and hole are situated on the same molecule. Consequently, the lowest singlet state possesses a charge transfer exciton character, and the lowest triplet exciton exhibits a Frenkel exciton character. The contact with the silicon substrate, as simulated by SSM, only slightly increases the charge transfer character of the singlet exciton by reducing the binding energy. 

The situation dramatically changes in the case of DSM. Taking into account the hybridization of the orbitals at the interface significantly affects the singlet exciton, while the triplet exciton still remains almost unchanged. In the former case, the crystalline symmetry of the silicon 1x2 reconstructed surface is imposed on the exciton, which results in an asymmetry of electron distribution relative to the center of the supercell where the hole is located. The side view of the electron density, shown in Fig. \ref{fig:wf1} a, reveals that the singlet exciton has a very pronounced inter-layer charge transfer character, with the election delocalized over the interface. Integrating this density over the region that confines the contact tetracene layer results in a value of 0.22 for the singlet exciton and 1.0 for the triplet exciton. Note, that the value 0.22 is the maximal value observed for all singlet excitons in the contact tetracene layer. 

The SF rate can be computed using the Fermi's Golden Rule \cite{Zaykov2019}: 
\begin{equation}
    W = 2 \pi \hbar^{-1} \left| \langle  S_1 \vert H_{int}  \vert T_1 T_1\rangle \right|^2 \rho(E)
    \label{w1}
\end{equation}
where $H_{int}$ is the interaction Hamiltonian and $\rho(E)$ is the density of states at the energy of the triplets. Next, we employ the following expansion of unity $1 = \sum_{\mathbf{r}} \vert \mathbf{r} \rangle \langle \mathbf{r} \vert$, 
where $\mathbf{r}$ is the radius-vector, and insert it from both the left and right sides of the interaction operator that leads to the following expression:
\begin{equation}
    W = 2 \pi \hbar^{-1} \rho(E)  \sum_{\mathbf{r}, \mathbf{r}'}  \left| \langle  S_1 \vert \mathbf{r} \rangle \langle \mathbf{r} \vert H_{int} \vert \mathbf{r}' \rangle \langle  \mathbf{r}'  \vert T_1 T_1\rangle \right|^2 
    \label{w3}
\end{equation}
In order to get a rough estimate of how the substrate affects the SF effect, we make the assumption that the interaction Hamiltonian decays rapidly with distance and use the approximation $H_{int}=\alpha \delta(\mathbf{r}-\mathbf{r}')$, which enables us to factor it out of the integral:
\begin{equation}
    W = 2 \pi \hbar^{-1} \vert \alpha  \vert^2  \rho(E) \sum_{\mathbf{r}} \vert \langle  S_1 \vert \mathbf{r} \rangle  \langle \mathbf{r} \vert T_1 T_1\rangle \vert^2 
    \label{w4}
\end{equation}

Assuming that the states $T_1$ are entirely confined within tetracene and considering that $|\langle \mathbf{r} | T_1 T_1 \rangle|$ is bounded from above by one, the upper bound for $W$ is:

\begin{equation}
    W  \sim 2 \pi \hbar^{-1} \vert \alpha  \vert^2  \rho(E) \sum_{\mathbf{r}_{Tc}} \vert \langle  S_1 \vert \mathbf{r} \rangle \vert^2 
    \label{w4}
\end{equation}
where $\mathbf{r}_{Tc}$  is constrained within the tetracene molecular layer. Both in the bulk tetracene and in an isolated tetracene slab, the value of $\sum_{\mathbf{r}_{Tc}} | \langle S_1 | \mathbf{r} \rangle |^2$ is close to one. This is because the state $S_1$ is entirely localized within the same layer as the states $T_1$. In the case of the contact with a silicon substrate, this value has been estimated to be approximately 0.22, as discussed earlier. Consequently, the SF rate is approximately 4.5 smaller, considering solely the overlap between the initial and product states.

Another, more dramatic implication for the SF effect in the presence of the substrate is the existence of an alternative relaxation path through the thermalization of the exciton in silicon. When the singlet state $S_1$ is hybridized with the continuum of unbound excitonic states in silicon, it can quickly undergo thermalization, resulting in a low-energy interlayer charge transfer exciton or Wannier-Mott exciton in silicon. In such a scenario, the efficient singlet-triplet down-conversion of the exciton is replaced by thermal losses. This process is facilitated by photon-assisted dephasing, which can take several femtoseconds in inorganic semiconductors at room temperature \cite{haug2009quantum}. Consequently, the singlet state $S_1$ has a shorter lifetime when it is in contact with the silicon substrate.

Thus, our findings indicate that the hybridization with unbound excitonic states in silicon reduces the probability of the SF process in the contact tetracene layer depicted by the equation \eqref{chain}. In the contact molecular layer, the dissociation of the singlet exciton after photo-excitation is more probable instead:

\begin{equation}
S_0S_0 \xrightarrow{h\nu} S_0S_1 \rightarrow S_0 + h_{Tc}^+ + e_{Si}^-    
\end{equation}

The triplet excitons are relatively less influenced by the interface and maintain long lifetimes. However, they cannot be generated through SF within the contacting layer. Instead, they are produced through the SF effect in neighboring layers and subsequently migrate to the contact layer due to their diffusion:

\begin{equation}
S_0^*S_0^* \xrightarrow{h\nu} S_0^*S_1^*  \xrightarrow{k_{fis}} T_1^* T_1^* \xrightarrow{k_{dif}} T_1 T_1 \xrightarrow{k_{tr}} S_0 + 2h_{Si}^+ + 2e_{Si}^-    
\end{equation}
where the asterisk denotes states in molecular layers that are not directly adjacent to the silicon surface and $k_{dif}$ is the diffusion rate.

Note, that SF in the contact tetracene layer can be restored by introducing a dielectric spacer between the substrate and tetracene \cite{Einzinger2019}.

\section{Conclusions}

In this work, we have demonstrated that when the singlet exciton in tetracene comes into contact with the clean 1x2 reconstructed Si(100) surface, it undergoes hybridization with the unbound excitonic states in silicon. As a result, the singlet exciton state $S_1$ exhibits a pronounced interlayer charge transfer character with delocalization of the exciton across the interface. For the singlet exciton, the maximum probability of both the electron and hole being located within the contacting tetracene layer does not exceed 0.22. Unlike the singlet exciton, the triplet exciton remains completely localized within the tetracene layer. This is a consequence of two key factors. First, the energy levels of the triplet exciton are positioned within the fundamental band gap and the triplet wave function is more localized. In contrast, the energy levels of the singlet excitons are embedded within the continuum of delocalized excitonic states of silicon and the singlet exciton has a more pronounced charge transfer character. Due to hybridization with silicon, the energy levels of the singlet exciton become broader.

The weak localization of the singlet exciton results in a smaller overlap with the product triplet states for the SF effect. The exciton hybridized with the unbound excitons in silicon is unstable, and its lifetime is significantly shorter compared to the singlet fission rate $k_{fis}$. This reduces the probability of SF for the contact tetracene layer. Furthermore, this hybridization leads to the emergence of an alternative relaxation pathway involving the thermalization process in silicon.

The presence of the silicon substrate significantly increases the singlet-triplet gap for the triplet excitons in the contact layer, by 144 meV, compared to the case of the isolated slab or the SSM model. Therefore, this deviation is primarily attributed to the hybridization with the unbound excitonic states in silicon leading to changes in the exchange energy. Our results indicate that the dynamic dielectric screening due to the substrate does not affect the singlet-triplet gap but does alter the exciton binding energies. This effect cannot be captured by models based on the "Add-chi" or dielectric embedding techniques. Nevertheless, these techniques are still valuable for predicting the impact of dielectric screening induced by the substrate on excitons, allowing us to isolate and study this specific effect independently from other factors.

\section{Acknowledgements}
The authors acknowledge the support of the Australian Research Council Center of Excellence in Exciton Science through grant CE170100026. Work at the Molecular Foundry was supported by the Office of Science, Office of Basic Energy Sciences of the U.S. Department of Energy under Contract No. DE-AC02-05CH11231. This project was undertaken with the assistance of resources and services from the National Computational Infrastructure, which is supported by the Australian Government.

\bibliography{bib}

\begin{figure}[t]
    \subfigure[]{\includegraphics[height=0.3\linewidth]{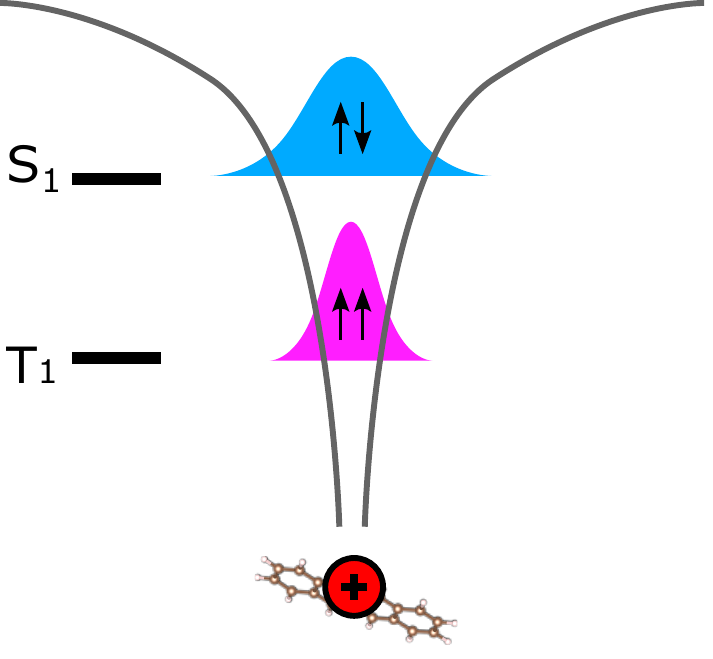}}
    \subfigure[]{\includegraphics[height=0.3\linewidth]{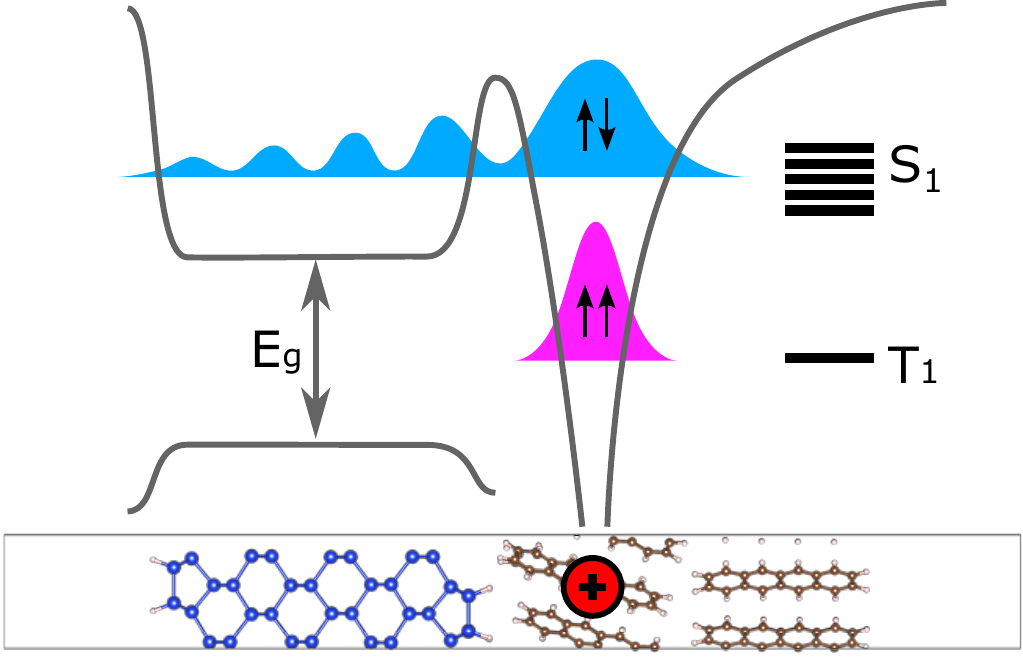}}
    \caption{TOC image}
    \label{fig:leaking_singlet}
\end{figure}

\end{document}